\documentclass{paper}

\preprintnumber{TTK-11-15\\SFB/CPP-11-23\\TTP11-15}

\title{Electroweak Precision Observables\\
  in a Fourth Generation Model\\
  with General Flavour Structure}

\author{%
  P.\ Gonz\'alez%
  \ref{inst:aachen}\email{gonzalez@physik.rwth-aachen.de}
  \and J.\ Rohrwild%
  \ref{inst:aachen}\email{rohrwild@physik.rwth-aachen.de}
  \and M.\ Wiebusch%
  \ref{inst:kit}\email{martin.wiebusch@kit.edu}%
}

\institutes{%
  \institute{inst:aachen}{%
    Institute for Theoretical Particle Physics and Cosmology,\\ 
    RWTH Aachen, D-52056 Aachen, Germany}%
  \institute{inst:kit}{Institute for Theoretical Particle Physics,\\
    Karlsruhe Institute of Technology (KIT), D-76128 Karlsruhe, Germany}%
}

\abstract{We calculate the contributions to electroweak precision observables
  (EWPOs) due to a fourth generation of fermions with the most general
  \mbox{(quark-)} flavour structure (but assuming Dirac neutrinos and a trivial
  flavour structure in the lepton sector). The new-physics contributions to the
  EWPOs are calculated at one-loop order using automated tools
  (FeynArts/FormCalc). No further approximations are made in our calculation. We
  discuss the size of non-oblique contributions arising from
  $Z$--quark--anti-quark vertex corrections and the dependence of the EWPOs on
  all CKM mixing angles involving the fourth generation. We find that the
  electroweak precision observables are sensitive to two of the
  fourth-generation mixing angles and that the corresponding constraints on
  these angles are competitive with those obtained from flavour physics. For
  non-trivial $4\times 4$ flavour structures, the non-oblique contributions lead
  to relative corrections of several permille and should be included in a global
  fit.}

\newcommand{\code}[1]{\texttt{#1}}
\newcommand{\Acal}{\mathcal{A}}
\newcommand{\Rcal}{\mathcal{R}}

\begin{document}
\maketitlepage
%
%
%%%%%%%%%%%%%%%%%%%%%%%%%%%%%%%%%%%%%%%%%%%%%%%%%%%%%%%%%%%%%%%%%%%%%%%%%%%%%%%%
\section{Introduction}\label{sec:intro}
%%%%%%%%%%%%%%%%%%%%%%%%%%%%%%%%%%%%%%%%%%%%%%%%%%%%%%%%%%%%%%%%%%%%%%%%%%%%%%%%
%
%
With the advent of LHC data the first direct tests for many models of new
physics are within reach. Among the conceptionally simplest extensions of the
Standard Model (SM3) are those which only add a minimal set of fermions to the
SM particle content. This class encompasses both the additional vector-like
quarks ~\cite{Branco1986738,Frampton:1999xi,delAguila:2000rc, Barenboim:2001fd} 
and the fourth-generation scenario (SM4).

The SM4 was fairly popular in the 1980s until electroweak precision observables
seemed to rule it out.  In the last years models with an additional fourth
generation experienced a renaissance as new analyses, e.g.~\cite{Holdom:1996bn,
  Maltoni:1999ta, He:2001tp, Novikov:2001md, Novikov:2002tk, Kribs:2007nz,
  Novikov:2009kc, Holdom:2009rf}, somewhat relaxed the electroweak
tensions. This realisation also prompted numerous studies of the non-trivial
flavour structure of the SM4~\cite{Alwall:2006bx, Bobrowski:2009ng,
  Chanowitz:2009mz,Hou:2010wf,Eberhardt:2010bm}, as well as searches for
specific signatures in new physics observables \cite{Hou:2006mx, Soni:2008bc,
  Soni:2010xh, Buras:2010pi, Buras:2010cp, Lacker:2010zz, Erler:2010sk}.
Currently, the progress of the LHC provides first direct constraints on new
SM-like heavy quarks. CMS gives lower mass limits of \unit{450}{GeV} for a $t'$
decaying into $bW$ \cite{CMS-PAS-EXO-11-051} and \unit{495}{GeV} for a $b'$
decaying into $tW$ \cite{CMS-PAS-EXO-11-036}.  Further constraints come from
direct Higgs searches at the LHC as the presence of additional heavy quarks
would dramatically increase the Higgs production cross section
\cite{Ruan:2011qg,Gunion:2011ww}. However, this constraint can be avoided if
the fourth generation neutrino is light enough to allow for an additional
invisible decay channel of the Higgs boson \cite{Belotsky:2002ym,
  Rozanov:2010xi, Keung:2011zc, Cetin:2011fp, Carpenter:2011wb}.
  
Recently, some effort has been directed towards providing an actual \emph{fit}
of the parameters of the model. One of the first attempts in this direction
primarily used the electroweak precision observables and restricted itself to
only one CKM parameter \cite{Chanowitz:2010bm}; still non-trivial correlations
were found, for example, between the Higgs mass and the new mixing angle.  More
recent studies seek to constrain \cite{Nandi:2010zx} or even determine
\cite{Alok:2010zj} the full $4\times 4$ CKM matrix.  In this case the main
challenge is the fact that, if one allows for a generic CKM structure, the
flavour and electroweak sector are intertwined and have to be treated
simultaneously.

Usually the effects of new physics in the electroweak sector are parametrised
by the \emph{oblique electroweak parameters} $S$, $T$ and $U$, as introduced by
Peskin and Takeuchi \cite{Peskin:1991sw,Maksymyk:1993zm}.  These allow for
fairly simple and straightforward estimates of new physics contribution to
electroweak observables.  However, the validity of this parametrisation relies
on certain assumptions about the new physics model, which are, in principle, no
longer satisfied in an SM4 with the most general flavour structure. Expressions
for the leading non-oblique contributions were given in \cite{Chanowitz:2009mz}
for the special case where only the third and fourth generation quarks are
allowed to mix.

In this letter we discuss the contributions to electroweak precision observables
\mbox{(EWPOs)} due to a fourth generation with general $4\times 4$ flavour
mixing.  For the sake of simplicity we assume Dirac neutrinos and a trivial
flavour structure in the lepton sector.  Our method for the computation of EWPOs
in the SM4 uses a high level of automatisation and can easily be used for other
models. We break up the observables into `SM3 parts' and `new physics
contributions'. The former can be computed with well-established programs like
\code{ZFITTER} \cite{Bardin:1999yd, Arbuzov:2005ma}. The latter we calculate to
one-loop order using \code{FeynArts/FormCalc} \cite{Hahn:2006qw, Hahn:2000kx,
  Hahn:1998yk}. Unlike most other literature on EWPOs in the SM4, no further
approximations are made in this paper. Our calculation thus includes all
one-loop non-oblique contributions, i.e.\ those which are not captured by the
$S$, $T$ and $U$ parameters.  We discuss the importance of the non-oblique
contributions and the impact of flavour mixing between the fourth and the first
three generation in several SM4 scenarios.

In section~\ref{sec:oblique} we briefly review the oblique parameters and their
range of applicability. In section~\ref{sec:Zqq} we introduce our notations for
the SM4 parameters and explain our method for calculating the corrections to the
EWPOs. In section~\ref{sec:GF} we describe our treatment of the Fermi constant
$G_F$, which is an observable and not a parameter in our analysis. Our numerical
results are presented in section~\ref{sec:results}. We find that the EWPOs are
equally sensitive to all three fourth-generation mixing angles and that the
corresponding constraints on these angles are competitive with those obtained
from flavour physics. For non-trivial $4\times 4$ flavour structures, the
non-oblique contributions lead to relative corrections of up to one permille for
the hadronic $Z$ width $\Gamma_\text{had}$ and of several permille for the
hadronic $Z\to b\bar b$ branching ratio $R_b$. A simultaneous fit of the SM4
masses, couplings and CKM matrix should therefore take into account all six SM4
CKM mixing angles and the non-oblique corrections to the EWPOs. We conclude in
section~\ref{sec:conclusions}.
%
%
%%%%%%%%%%%%%%%%%%%%%%%%%%%%%%%%%%%%%%%%%%%%%%%%%%%%%%%%%%%%%%%%%%%%%%%%%%%%%%%%
\section{Oblique Corrections and Electroweak Observables}\label{sec:oblique}
%%%%%%%%%%%%%%%%%%%%%%%%%%%%%%%%%%%%%%%%%%%%%%%%%%%%%%%%%%%%%%%%%%%%%%%%%%%%%%%%
%
%
The constraints imposed on new physics by EWPOs measured at LEP have already
been discussed extensively in the literature. In 1992 Peskin and Takeuchi
presented a model-independent way of parametrising the new physics contributions
to the $Z$ pole observables \cite{Peskin:1991sw}. Their analysis was based on
three assumptions:
\begin{enumerate}
\item The electroweak gauge group of the new-physics model is $SU(2)_L\times
  U(1)_Y$.
\item The new-physics couplings to light fermions (i.e.\ all SM3 fermions except
  the top-quark) are negligible.
\item The scale of new physics is much larger than the electroweak scale.
\end{enumerate}
The first assumption forbids the existence of additional gauge bosons coupling
directly to leptons. The second assumption guarantees that there are no
additional vertex or box-diagrams contributing to the Drell-Yan process. Thus,
the only way the new physics contribute to the $Z$ pole observables is through
the renormalisation of weak gauge boson wave functions, the electric charge or
the Weinberg angle. The third assumption is needed to justify a step in the
discussion in \cite{Peskin:1991sw}, where the gauge boson self-energies are
expanded to first order around $q^2=0$ ($q$ being the momentum flowing through
the self-energy graphs). In practice, it is usually sufficient to require that
new particles coupling directly to weak gauge bosons are heavier than the $Z$
boson.

In SM extensions that satisfy the criteria above, the new physics contributions
to the $Z$ pole observables can be expressed in terms of the \emph{oblique
  electroweak parameters} $S$, $T$ and $U$ which were defined in
\cite{Peskin:1991sw} and represent different linear combinations
of gauge boson self-energies and their derivatives. On the experimental side,
the values of $S$, $T$ and $U$ can then be determined from data by performing a
global fit of $S$, $T$, $U$ and the SM3 parameters to the $Z$ pole and possibly
other low-energy observables. (See \cite{:2010vi} for a recent analysis of this
type.) On the theoretical side one can test to what extent a given model of new
physics agrees with low-energy observables by computing $S$, $T$ and $U$ in this
model and comparing the results with the best-fit values.

This method of testing an SM extension against constraints from low-energy
experiments is very convenient since it only requires the computation of three
quantities. It has been applied to a number of models including the SM4
\cite{Kribs:2007nz}. One should, however, keep in mind that the validity of this
method depends on the validity of the assumptions listed above. In the SM4 the
second assumption is no longer valid if the fourth generation quarks are allowed
to mix with the quarks of the first three generations. Furthermore, if one
allows for a fourth-generation neutrino with a mass just above $m_Z/2$ the third
assumption is no longer satisfied.\footnote{If the first two assumptions are
  still valid, the case of new physics near the electroweak scale can be handled
  by introducing three additional oblique parameters. This was discussed in
  \cite{Maksymyk:1993zm}.} Hence, the validity of the ``oblique method'' must be
checked explicitly if one attempts to constrain the new mixing angles of the SM4
CKM matrix.
%
%
%
%%%%%%%%%%%%%%%%%%%%%%%%%%%%%%%%%%%%%%%%%%%%%%%%%%%%%%%%%%%%%%%%%%%%%%%%%%%%%%%%
\section{The \texorpdfstring{$\boldsymbol{Zq\bar q}$}{Z-q-qbar}
  Vertex in the SM4}\label{sec:Zqq}
%%%%%%%%%%%%%%%%%%%%%%%%%%%%%%%%%%%%%%%%%%%%%%%%%%%%%%%%%%%%%%%%%%%%%%%%%%%%%%%%
%
%
\begin{table}
  \begin{center}
    \begin{tabular}{cr@{.}l@{${}\pm{}$}r@{.}lr@{.}l@{${}\pm{}$}r@{.}l}
      \hline\hline 
      & \multicolumn{4}{c}{experiment} & \multicolumn{4}{c}{theory (SM3)} \\
      \hline
      $\Gamma_\text{had}$ [GeV]& 1&7444 & 0&002 &  1&7418  & 0&0009  \\
      $R_b$        & 0&21629 & 0&00066          &  0&21578 & 0&00005 \\
      $A_{FB}^b$   & 0&0992  & 0&0016           &  0&1034  & 0&0007  \\
      $\Acal_b$    & 0&923   & 0&020            &  0&9348  & 0&0001  \\
      $\Acal_e$    & 0&15138 & 0&00216          &  0&1475  & 0&0010  \\
      $M_W$ [GeV]  & 80&420  & 0&031            &  80&384  & 0&014   \\
      \hline\hline
    \end{tabular}
  \end{center}
  \caption{Experimental results and Standard Model predictions for selected
    electroweak observables.  All numbers were taken from
    \cite{Nakamura:2010zzi}.}
  \label{tab:EWOdata} 
\end{table}
The properties of the $Z$ boson and its couplings to fermions have been measured
at LEP 1 with a very high accuracy. Table~\ref{tab:EWOdata} shows the
experimental values and accuracies for a selection of $Z$-pole observables as
well as their theoretical predictions within the SM3.  The observables are: the
partial width for $Z\to\text{hadrons}$ ($\Gamma_\text{had}$), the hadronic
branching fraction for $Z\to b\bar b$ ($R_b$), the forward-backward asymmetry
for $Z\to b\bar b$ ($A^b_\text{FB}$) and the mass of the $W$ ($M_W$).  In the
$Z$-pole approximation, the forward-backward asymmetry can be written as
$\frac34\Acal_e\Acal_b$, where the quantities $\Acal_e$ and $\Acal_b$ only
depend on the $Ze^+e^-$ and $Zb\bar b$ couplings, respectively.  The relative
precision of $\Gamma_\text{had}$ is approximately $0.1\%$ and $R_b$ is known
to an accuracy of $0.3\%$. The measured value of $A_\text{FB}^b$ deviates from
its SM3 prediction by more than two standard deviations. The discrepancy
originates mainly from the factor $\Acal_e$.  Oblique corrections due to a
fourth generation of fermions affect all $Z$-pole observables, but only
observables related to the $Z$--quark--anti-quark vertex are subject to
non-oblique corrections; of the observables from table~\ref{tab:EWOdata}, only
$\Gamma_\text{had}$, $R_b$ and $\Acal_b$ receive non-oblique contributions.  Our
discussion will therefore mainly focus on these quantities.\footnote{The
  branching fraction $R_c$ and asymmetry factor $\Acal_c$ for the charm quark
  also receive non-oblique corrections, but these observables are less
  constraining due to their larger experimental error.}

Within the SM3 the couplings of $Z$ bosons to quarks have been studied in great
detail.  Electroweak and QCD corrections to the gauge boson self-energies and
the $Z$--quark--anti-quark vertex have been calculated at two-loop order
\cite{Degrassi:1996mg, Degrassi:1999jd, Freitas:2000gg, Freitas:2002ja,
  Awramik:2003ee, Awramik:2002vu, Chetyrkin:1995js, Chetyrkin:1996cf,
  Faisst:2003px, Awramik:2004ge, Barbieri:1992nz, Barbieri:1992dq,
  Fleischer:1994cb, Degrassi:1996ps, Djouadi:1987gn, Kniehl:1989yc,
  Halzen:1990je, Kniehl:1991gu, Kniehl:1992dx, Djouadi:1993ss}
%% \cite{CERN9503, Bardin:1998nm, Bardin:1999gt, Degrassi:1999jd, Gorishnii:1990vf,
%%   Surguladze:1990tg, Chetyrkin:1996ez, Chetyrkin:1990kr, Chetyrkin:1996hm,
%%   Kniehl:1989qu, Larin:1994va, Chetyrkin:1993ug, Chetyrkin:1996ia,
%%   Akhundov:1985fc, Beenakker:1988pv, Bernabeu:1987me, Lynn:1990hd,
%%   Barbieri:1992dq, Fleischer:1993ub, Fleischer:1992fq, Buchalla:1992zm,
%%   Degrassi:1993ij, Chetyrkin:1993jp, Kwiatkowski:1994ig, Peris:1995pm,
%%   Kataev:1992dg, Czarnecki:1996ei, Fleischer:1999iq, Harlander:1997zb}
and the results have been implemented in public codes such as \code{TOPAZ0}
\cite{Montagna:1998kp} or \code{ZFITTER} \cite{Bardin:1999yd, Arbuzov:2005ma}.
Radiative corrections to the partial widths are of the order of 0.1\% (QED) and
4\% (QCD).  To match the experimental accuracy of the $Z$-pole observables they
must therefore be included in theoretical calculations. In this section we
explain how predictions for the $Z$ pole observables within the SM4 can be
calculated at the required level of accuracy without the need to re-visit the
SM3 calculations.

Before we begin, let us briefly explain our notations for the SM3 and SM4
parameters. For the SM3 CKM matrix we use the \emph{standard parametrisation}.
In this parametrisation the independent parameters are the three mixing angles
$\theta_{12}$, $\theta_{13}$ and $\theta_{23}$ and one complex phase
$\delta_{13}$. The explicit form of the SM3 CKM matrix in terms of the phase and
mixing angles is given in appendix~\ref{sec:CKM}.

In the SM4 the CKM matrix is a unitary $4\times 4$ matrix. After absorbing
unphysical complex phases into the definitions of the quark fields, its
parametrisation requires only three additional mixing angles $\theta_{14}$,
$\theta_{24}$ and $\theta_{34}$ and two additional complex phases $\delta_{14}$
and $\delta_{24}$. The explicit form of the SM4 CKM matrix is also given in
appendix~\ref{sec:CKM}. For the discussion below it is only important to know
that for $\theta_{14}=\theta_{24}=\theta_{34}=\delta_{14}=\delta_{24}=0$ the SM4
CKM matrix assumes a block-diagonal form with the SM3 CKM matrix in the first
$3\times 3$ block and a one in the last block.

To distinguish the phase $\delta_{13}$ and the mixing angles $\theta_{12}$,
$\theta_{13}$ and $\theta_{23}$ of the SM4 CKM matrix from their SM3
counterparts we will use superscripts `SM4' and `SM3', respectively. The same
applies to other parameters like $m_H$ or $M_W$, which exist in both models.
We will also use the shorthands $s_{ij}$ and $c_{ij}$ for the sines and cosines
of the mixing angles $\theta_{ij}$. Finally, we denote the lepton, neutrino, up
and down-type quark of the fourth generation as $\ell_4$, $\nu_4$, $t'$ and
$b'$, respectively. Their masses $m_{\ell_4}$, $m_{\nu_4}$, $m_{t'}$ and
$m_{b'}$ are independent parameters of the SM4.

Let us now proceed with the discussion of higher order corrections to the
$Zq\bar q$ vertex. In the limit of vanishing external quark masses $m_q$, the
on-shell $Zq\bar q$ vertex function only contains two Lorentz structures:
\begin{equation}
  \Gamma^q_\mu = ie\gamma_\mu[F^q_V - F^q_A\gamma_5]
  \eqpunct.
\end{equation}
Here and in the following, $q=u,d,s,c,b$ denotes the quark flavour. The form
factors $F^q_V$ and $F^q_A$ depend on the quark flavour, the external masses and
the parameters of the model under consideration (SM3 or SM4). Following the
discussion in \cite{CERN9503}, we express QCD and QED radiative corrections to
$F^q_V$ and $F^q_A$ in terms of \emph{radiator functions} $\Rcal^q_V$ and
$\Rcal^q_A$ and write
\begin{equation}
  F^q_V = g^q_V\sqrt{\Rcal^q_V}
  \eqsep,\eqsep
  F^q_A = g^q_A\sqrt{\Rcal^q_A}
  \eqsep.
\end{equation}
In doing this, we neglect the non-factorisable contributions
\cite{Czarnecki:1996ei, Harlander:1997zb}, whose effect is below the permille
level. The \emph{effective couplings} $g^q_V$ and $g^q_A$ now only contain
infrared finite contributions. At leading order $\Rcal^q_V=\Rcal^q_A=1$ and
$g^q_V$ and $g^q_A$ are the tree-level vector and axial couplings of the $Z$
boson.

In this paper we are interested in the difference between predictions for $Z$
pole observables within the SM3 and SM4. For this purpose we denote, for any
quantity $X$, the \emph{new physics correction} by
\begin{equation}
  \delta X = X^\text{SM4} - X^\text{SM3}
  \eqsep,
\end{equation}
where the superscripts `SM4' and `SM3' indicate that $X$ is evaluated with a
given set of SM4 or SM3 parameters, respectively. In principle, the two sets of
parameters can be completely unrelated. It is, however, extremely convenient to
use the same values of $M_Z$, $M_W$, $m_t$, $\alpha$ and $\alpha_s$ in both
sets.\footnote{These are independent SM3 input parameters in the \emph{on-shell}
  renormalisation scheme ~\cite{Bohm:1986rj}, which is the scheme we used in our
  calculations.} In this case, $\delta \Rcal^q_V = \delta \Rcal^q_A = 0$ and the
new physics corrections to any $Z$ pole observable can be obtained by only
computing the infrared finite quantities $\delta g^q_V$ and $\delta g^q_A$.  The
form factors $F_V^{q,\text{SM4}}$ and $F_A^{q,\text{SM4}}$ (and thus for the
$Z$-pole observables within the SM4) may then be calculated by scaling the
corresponding SM3 form factors with the ratios
$g_V^{q,\text{SM4}}/g_V^{q,\text{SM3}}$ and
$g_A^{q,\text{SM4}}/g_A^{q,\text{SM3}}$, respectively. This way, factorisable
QCD and QED corrections are included in $F_V^\text{SM4}$ and $F_A^\text{SM4}$ if
they were included in the SM3 `reference values' $F_V^\text{SM3}$ and
$F_A^\text{SM3}$.  As we will see below, the ratios $\delta g^q_V/g^{q(0)}_V$
and $\delta g^q_A/g^{q(0)}_A$ (with $g^{q(0)}_V$ and $g^{q(0)}_A$ being the
tree-level couplings) are typically below 1\%. Thus, the approximation
\begin{equation}
    F_V^{q,\text{SM4}}
  \approx F_V^{q,\text{SM3}}\left(1+\frac{\delta g^q_V}{g_V^{q,(0)}}\right)
  \eqsep,\eqsep
    F_A^{q,\text{SM4}}
  \approx F_A^{q,\text{SM3}}\left(1+\frac{\delta g^q_A}{g_A^{q,(0)}}\right)
\end{equation}
is generally valid with a relative precision of the order of $10^{-4}$.

The difference between $\Rcal^q_V$ and $\Rcal^q_A$ is of the order of a few
percent \cite{CERN9503}. Thus, to estimate the size of the new physics
contributions to the EWPOs we use the approximation
\begin{equation}
  \Rcal^q_V \approx \Rcal^q_A \equiv \Rcal^q
\end{equation}
and obtain
\begin{equation}
  \delta\Gamma^q_\mu = ie\Rcal^q\gamma_\mu[\delta g^q_V - \delta g^q_A\gamma_5]
  \eqpunct.
\end{equation}
The hadronic $Z$ partial widths and asymmetries are then given by
\begin{equation}
  \Gamma(Z\to q\bar q) = \alpha M_Z\Rcal^q[|g_V^q|^2 + |g_A^q|^2]
  \quad,\quad
  \Acal_q = \frac{2\re g^q_V\re g^q_A}{(\re g^q_V)^2+(\re g^q_A)^2}
  \eqsep.
\end{equation}
The new physics corrections to these quantities are readily obtained by
expanding the effective couplings to first order in $\delta g^q_V$ and $\delta
g^q_A$:
\begin{subequations}
\begin{align}
  \frac{\delta\Gamma(Z\to q\bar q)}{\Gamma^\text{SM3}(Z\to q\bar q)}
  &= 2\,\frac{  \re[g^{q(\text{SM3})*}_V\delta g^q_V]
              + \re[g^{q(\text{SM3})*}_A\re\delta g^q_A]}%
             {|g^{q(\text{SM3})}_V|^2 + |g^{q(\text{SM3})}_A|^2}
  \nonumber\\
  &\approx 2\,\frac{  g^{q(0)}_V\re\delta g^q_V
                    + g^{q(0)}_A\re\delta g^q_A}%
                   {(g^{q(0)}_V)^2 + (g^{q(0)}_A)^2}
  \eqsep,\\
  \frac{\delta\Acal_q}{\Acal_q^\text{SM3}}
  &=   \frac{\re\delta g^q_V}{\re g^{q(\text{SM3})}_V}
     + \frac{\re\delta g^q_A}{\re g^{q(\text{SM3})}_A}
     - 2\,\frac{  \re g^{q(\text{SM3})}_V\re\delta g^q_V
                + \re g^{q(\text{SM3})}_A\re\delta g^q_A}%
               {(\re g^{q(\text{SM3})}_V)^2 + (\re g^{q(\text{SM3})}_A)^2}
  \nonumber\\
  &\approx   \frac{\re\delta g^q_V}{g^{q(0)}_V}
           + \frac{\re\delta g^q_A}{g^{q(0)}_A}
           - 2\,\frac{  g^{q(0)}_V\re\delta g^q_V
                      + g^{q(0)}_A\re\delta g^q_A}%
                     {(g^{q(0)}_V)^2 + (g^{q(0)}_A)^2}
  \eqsep.
\end{align}\label{eq:ratios}%
\end{subequations}
Note that, as a result of approximating $\Rcal^q_V \approx \Rcal^q_A$, the
radiator functions cancel in the ratios above. 

If mixing between the fourth generation quarks and the quarks of the first three
generations is neglected and the fourth-generation fermions are sufficiently
heavy, the new physics corrections can be expressed in terms of the oblique
electroweak parameters $S$, $T$ and $U$ \cite{Peskin:1991sw,Maksymyk:1993zm}.
In this case, the relations between $\delta g^q_V$, $\delta g^q_A$ and $S$, $T$
and $U$ are
\begin{subequations}
\begin{align}
  \delta g^{q,(1)}_V &= \frac{\alpha}{16 c_W s_W^3}\left[
      2 I_3^q S - 4 [(c_W^2-s_W^2) I_3^q + 2 s_W^2 Q^q] T
    - \Bigl(\frac{c_W^2-s_W^2}{s_W^2} I_3^q + 2 Q^q\Bigr) U\right]
  \eqsep,\\
  \delta g^{q,(1)}_A &= \frac{\alpha}{16 c_W s_W^3}\left[
    2 S - \frac{c_W^2-s_W^2}{s_W^2}(4 s_W^2 T + U)\right]I_3^q
  \eqsep,
\end{align}\label{eq:oblique}%
\end{subequations}
where $Q^q$ and $I_3^q$ are the electric charge and weak isospin of the quark
$q$ and $s_W$ and $c_W$ are the sine and cosine of the Weinberg angle, defined
by $s_W^2 = 1 - M_W^2/M_Z^2$. The superscript `$(1)$' denotes the (electroweak)
one-loop corrections to the effective couplings.\footnote{If $\alpha(M_Z)$,
  $M_Z$ and $M_W$ are the same in the SM3 and SM4, the new physics only enters
  at one-loop order. In the next section we deal with the case where a different
  value of $M_W$ is chosen in the two models.}

If the fourth generation quarks are allowed to mix with the quarks of the first
three generations one also needs to compute the vertex diagrams contributing to
$\delta g^{q,(1)}_V$ and $\delta g^{q,(1)}_A$. We used the
\code{FeynArts/FormCalc} package \cite{Hahn:2006qw, Hahn:2000kx, Hahn:1998yk} to
compute the required one-loop order diagrams. The renormalisation of the $Zq\bar
q$ vertex was done in the on-shell scheme.\footnote{See e.g.~\cite{Denner} for
  a detailed description.} Only diagrams involving $W$ bosons, charged Goldstone
bosons or Higgs bosons contribute to $\delta g^{q,(1)}_V$ and $\delta
g^{q,(1)}_A$, as long as $\alpha$, $\alpha_s$, $M_Z$, $M_W$ and $m_t$ are chosen
to be the same in the SM3 and SM4. The SM3 parameters and corresponding values
for $\Gamma(Z\to q\bar q)$ and $\Acal_q$ were taken from
\cite{Nakamura:2010zzi}. Specifically, we use
\begin{gather}
  1/\alpha(m_Z) = 128.892
  \quad,\quad
  \alpha_s(m_Z) = 0.1185
  \quad,\quad 
  M_Z = \unit{91.1875}{GeV}
  \eqpunct,\nonumber\\
  M_W^\text{SM3} = \unit{80.384}{GeV}
  \quad,\quad 
  m_t = \unit{173.2}{GeV}
  \quad,\quad 
  m_H^\text{SM3} = \unit{90}{GeV}
  \eqpunct,\nonumber\\
  \Gamma^\text{SM3}_\text{had} = \unit{1.7418}{GeV}
  \quad,\quad
  R_b^\text{SM3} = 0.21578
  \quad,\quad
  \Acal_b^\text{SM3} = 0.9348
  \eqpunct,\label{eq:SM3_parameters}
\end{gather}
where
\begin{equation}
  \Gamma_\text{had} = \sum_{q=u,d,s,c,b}\Gamma(Z\to q\bar q)
  \quad,\quad
  R_q = \frac{\Gamma(Z\to q\bar q)}{\Gamma_\text{had}}
  \eqpunct.
\end{equation}
The phase and mixing angles of the SM3 CKM matrix were also taken from
\cite{Nakamura:2010zzi}:
\begin{equation}\label{eq:SM3_angles}
  \theta^\text{SM3}_{12} = 0.2273
  \quad,\quad
  \theta^\text{SM3}_{13} = 0.003466
  \quad,\quad
  \theta^\text{SM3}_{23} = 0.04103
  \quad,\quad
  \delta^\text{SM3}_{13} = 1.2020
  \eqsep.
\end{equation}
Note that the numerical values for $\Gamma^\text{SM3}_\text{had}$ and
$R_b^\text{SM3}$ are for a fixed ``reference'' Higgs mass
$m_H^\text{SM3}=\unit{90}{GeV}$. In the SM4 the Higgs mass is treated as a free
parameter.
%
%
%
%%%%%%%%%%%%%%%%%%%%%%%%%%%%%%%%%%%%%%%%%%%%%%%%%%%%%%%%%%%%%%%%%%%%%%%%%%%%%%%%
\section{A Note on \texorpdfstring{$\boldsymbol{G_F}$}{GF}}\label{sec:GF}
%%%%%%%%%%%%%%%%%%%%%%%%%%%%%%%%%%%%%%%%%%%%%%%%%%%%%%%%%%%%%%%%%%%%%%%%%%%%%%%%
%
%
As mentioned above, we use in this work the on-shell renormalisation scheme
for the computation of new physics corrections. In this scheme, the quantities
$\alpha(M_Z)$, $M_Z$ and $M_W$ are independent parameters. This parametrisation
is very convenient for the computation of higher order corrections, but it has
its disadvantages if one wants to compare it with experimental data. The Fermi
constant $G_F$, which is determined from the muon lifetime, is a non-trivial
function of $\alpha(M_Z)$, $M_Z$, $M_W$ and the other model parameters. Since
$G_F$ is measured very accurately (namely, to a relative precision of $10^{-5}$)
it constrains the model to a non-trivial hyper-surface in its parameter
space. In other words, one parameter of the model is fixed by the requirement
that $G_F$ assumes its measured value. Typically, one adjusts the value of $M_W$
to obtain the correct value of $G_F$.

The relation between $G_F$ and $M_W$ is conventionally written as
\cite{Sirlin:1980nh}
\begin{equation}\label{eq:GF_constraint}
  G_F = \frac{\pi\alpha}{\sqrt2 s_W^2M_W^2}\frac1{1-\Delta r}
  \eqsep,
\end{equation}
where $\Delta r$ encodes higher order corrections and is, in general, a function
of all other parameters. New physics, like the existence of a fourth
generation of fermions, changes the function $\Delta r$. Denoting, as before,
the new physics correction to $\Delta r$ as $\delta\Delta r$ and writing the
solutions of \eqref{eq:GF_constraint} in the SM3 and SM4 as $M_W^\text{SM3}$
and $M_W^\text{SM4}\equiv M_W^\text{SM3}+\delta M_W$, respectively, we find
\begin{equation}
  \frac{\delta M_W}{M_W^\text{SM3}}
  = -\frac{s_W^2}{2(c_W^2-s_W^2)}\delta\Delta r
  \eqsep.
\end{equation}
However, since the parameters \eqref{eq:SM3_parameters} already satisfy the
$G_F$ constraint we have $M_W^\text{SM3}=M_W$ with $M_W$ from
\eqref{eq:SM3_parameters}. If the SM4 is to agree with the measured value of the
$W$ mass, the ratio $\delta M_W/M_W^\text{SM3}$ cannot be much larger than one
permille. Hence, the shift in $M_W$ is unimportant for the purpose of computing
$\delta\Delta r$ and the loop corrections to the effective couplings $g^q_V$ and
$g^q_A$. We can therefore safely use the value from \eqref{eq:SM3_parameters} in
these calculations. However, the change in $M_W$ also affects the tree-level
$Zq\bar q$ couplings $g_V^{q,(0)}$ and $g_A^{q,(0)}$, since $\sin^2\theta_W$ is
defined as $1-M_W^2/M_Z^2$ in the on-shell scheme. This effect is of the same
order as the loop corrections to the effective couplings and must be included
in the new-physics corrections $\delta g^q_V$ and $\delta g^q_A$. This can be
achieved by making the following substitution in \eqref{eq:ratios}:
\begin{equation}
    \delta g^q_{V,A}
  = g^{q,(0)}_{V,A}|_{M_W=M_W^\text{SM4}}
   -g^{q,(0)}_{V,A}|_{M_W=M_W^\text{SM3}}
   +\delta g^{q,(1)}_{V,A}|_{M_W=M_W^\text{SM3}}
  \eqsep,
\end{equation}
where $\delta g^{q,(1)}_{V,A}$ denotes the new-physics corrections of the
one-loop contributions to $g^q_{V,A}$.
%
%
%
%%%%%%%%%%%%%%%%%%%%%%%%%%%%%%%%%%%%%%%%%%%%%%%%%%%%%%%%%%%%%%%%%%%%%%%%%%%%%%%%
\section{Numerical Results}\label{sec:results}
%%%%%%%%%%%%%%%%%%%%%%%%%%%%%%%%%%%%%%%%%%%%%%%%%%%%%%%%%%%%%%%%%%%%%%%%%%%%%%%%
%
%
In this section we study the dependence of the effective couplings and the $Z$
pole observables on the masses and mixing angles of the fourth generation.
To this end we set
\begin{gather}
  \theta^\text{SM4}_{12}=\theta^\text{SM3}_{12}
  \eqsep,\eqsep
  \theta^\text{SM4}_{13}=\theta^\text{SM3}_{13}
  \eqsep,\eqsep
  \theta^\text{SM4}_{23}=\theta^\text{SM3}_{23}
  \eqsep,\eqsep
  \delta^\text{SM4}_{13}=\delta^\text{SM3}_{13}
  \eqsep,\nonumber\\
  \delta_{14} = \delta_{24} = 0
  \eqsep,\nonumber\\
  m_{\ell_4} = \unit{101}{GeV}
  \eqsep,\eqsep  
  m_{\nu_4} = \unit{50}{GeV}
  \eqsep,\eqsep  
  m_{t'} = m_{b'} = \unit{500}{GeV}
  \eqsep,\label{eq:first_scenario}
\end{gather}
with $\theta^\text{SM3}_{12}$, $\theta^\text{SM3}_{13}$,
$\theta^\text{SM3}_{23}$ and $\delta^\text{SM3}_{13}$ from
\eqref{eq:SM3_angles}. In this scenario the values of $m_{t'}$ and $m_{b'}$
evade the current CMS limits on fourth-generation quarks
\cite{CMS-PAS-EXO-11-051, CMS-PAS-EXO-11-036}. The constraints from direct Higgs
searches at the LHC are avoided by choosing a sufficiently small
fourth-generation neutrino mass, so that the Higgs boson can decay invisibly
into $\nu_4\bar\nu_4$ \cite{Belotsky:2002ym, Rozanov:2010xi, Keung:2011zc,
  Cetin:2011fp}.

The fourth-generation phases $\delta_{14}$ and $\delta_{24}$ have no noticeable
impact on the effective couplings $g^q_V$ and $g^q_A$ since they only depend on
moduli of the $V_\text{CKM4}$ matrix elements. We have checked explicitely that
the variations in the EWPOs are far below their experimental uncertainties when
the new $CP$ phases are varied between $0$ and $2\pi$.

The absolute values of the entries in the first two rows of the $3\times 3$ CKM
matrix are strongly constrained by various flavour-observables. In the context
of the SM4 these constraints were first discussed in \cite{Bobrowski:2009ng} and
it was found that scenarios are allowed where the mixing angles $\theta_{12}$,
$\theta_{13}$ and $\theta_{23}$ deviate substantially from their best-fit SM3
values. In \cite{Chanowitz:2009mz} it was then pointed out that electroweak
precision measurements rule out these large-mixing scenarios and the analysis
\cite{Bobrowski:2009ng} was updated in \cite{Eberhardt:2010bm}, where
electroweak precision constraints were implemented via the oblique parameters
$S$ and $T$. Here we are mainly interested in the difference between the oblique
and the full corrections to the EWPOs. To simplify the numerical discussion we
therefore keep the SM3 mixing angles fixed according to \eqref{eq:SM3_angles}
and only vary the new mixing angles.  For these we will use the following
independent limits from \cite{Eberhardt:2010bm}:
\begin{equation}
  |s_{14}| < 0.05
  \eqsep,\eqsep
  |s_{24}| < 0.1
  \eqsep,\eqsep
  |s_{34}| < 0.2
  \eqsep,
\end{equation}
with $s_{ij}\equiv\sin\theta_{ij}$. In a future combined fit of flavour and
electroweak precision observables all mixing angles should be varied
independently, and the non-oblique corrections will then also affect the
best-fit values of the SM3 mixing angles.

\begin{figure}
\centering\includegraphics{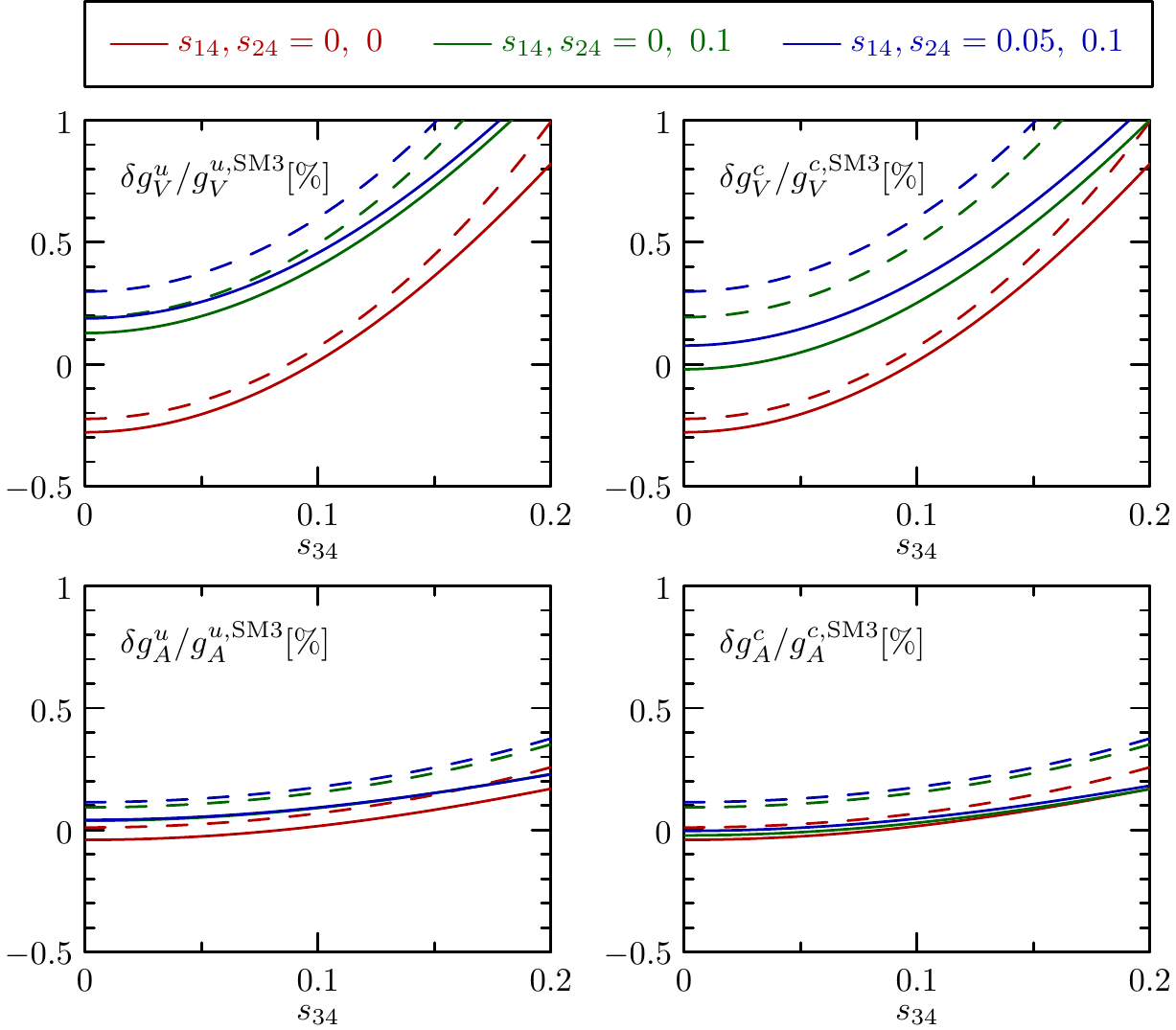}
\caption{The relative `new physics' corrections to the effective couplings
  $g^q_V$ and $g^q_A$ for up-type quarks ($q=u,c$) as functions of $s_{34}$ for
  the parameters \eqref{eq:first_scenario} and different combinations of
  $s_{14}$ and $s_{24}$. The dashed lines show the oblique corrections as
  computed from \eqref{eq:oblique}.}
\label{fig:guVguA-s14s24}
\end{figure}
\begin{figure}
\centering\includegraphics{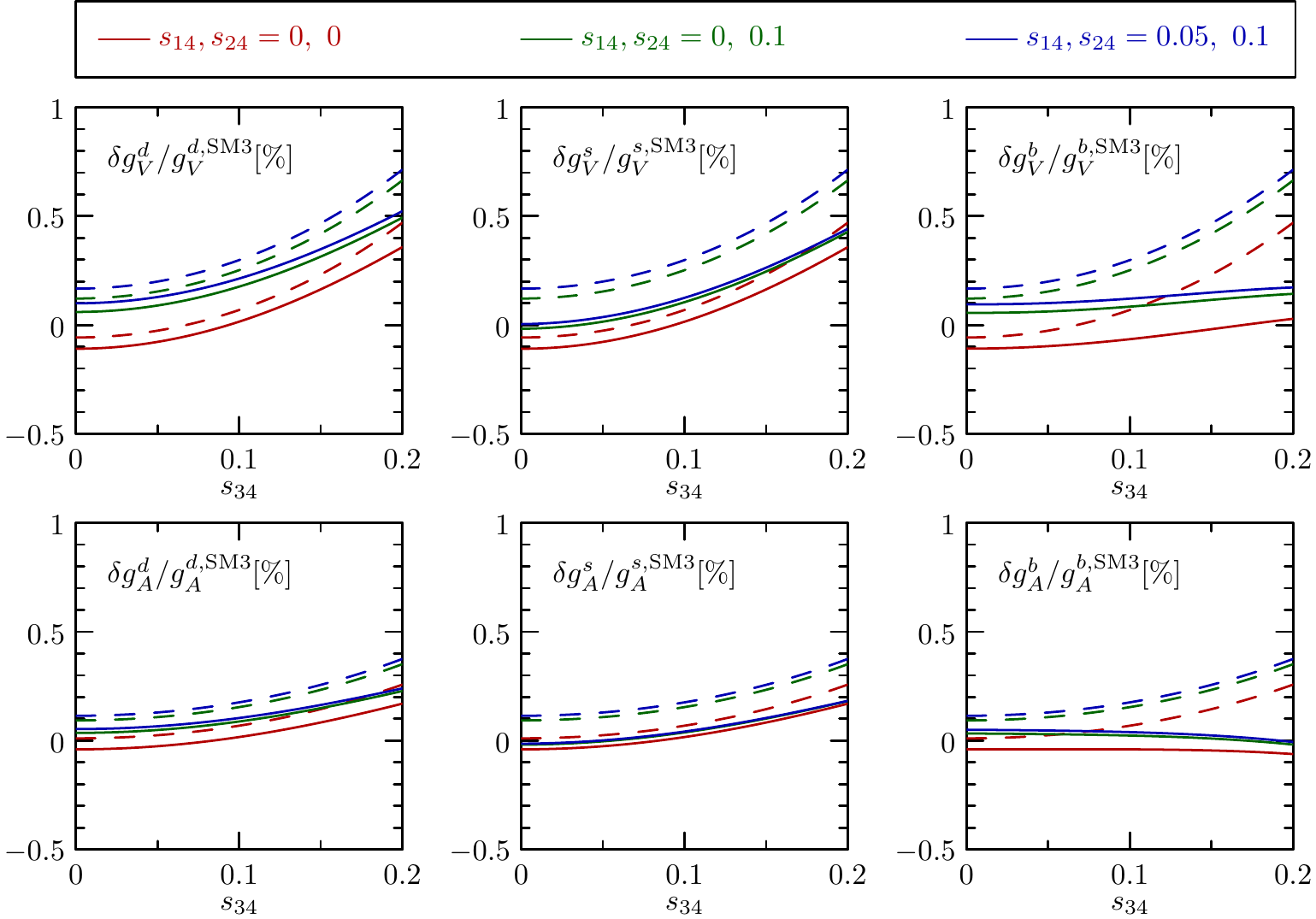}
\caption{The relative `new physics' corrections to the effective couplings
  $g^q_V$ and $g^q_A$ for down-type quarks ($q=d,s,b$) as functions of $s_{34}$
  for the parameters \eqref{eq:first_scenario} and different combinations of
  $s_{14}$ and $s_{24}$. The dashed lines show the oblique corrections as
  computed from \eqref{eq:oblique}.}
\label{fig:gdVgdA-s14s24}
\end{figure}
Figure \ref{fig:guVguA-s14s24} and \ref{fig:gdVgdA-s14s24} show the relative new
physics corrections to $g^q_V$ and $g^q_A$ for up-type and down-type quarks $q$,
respectively, as functions of $s_{34}$ and for different combinations of
$s_{14}$ and $s_{24}$. Varying $s_{24}$ $s_{34}$ in their allowed ranges leads
to effects above the permille level. Hence, one should expect that EWPOs
constrain $\theta_{24}$ and $\theta_{34}$ simultaneously. The contributions
which depend on $s_{14}$ and $s_{24}$ are positive and the effects are additive:
varying one mixing angle shifts the effective couplings by amounts which are
mostly independent of the other mixing angles. This can be explained by the fact
that the largest $s_{14}$ or $s_{24}$-dependent contributions come from terms
where one angle enters through a sine factor and the others through cosine
factors. The highest sensitivity to $s_{14}$ and $s_{24}$ is observed in the
vector couplings of up-type quarks ($g^u_V$ and $g^c_V$). Varying $s_{14}$,
$s_{24}$ and $s_{34}$ in the chosen ranges leads to corrections between $-0.3$
and $1.2\%$ for up-type quarks and between $-0.1$ and $0.5\%$ for down-type
quarks.

The dashed lines in Figures~\ref{fig:guVguA-s14s24} and \ref{fig:gdVgdA-s14s24}
contain only the oblique corrections while the solid lines are the result of our
``exact'' calculation. As explained earlier there are two reasons for their
discrepancy: the mixing between all four generations of quarks leads to vertex
corrections that are not included in the oblique parameters and the small mass
of the fourth-generation neutrino makes the expansions used in the derivation of
$S$, $T$ and $U$ invalid. We see that the non-oblique effects are at the level
of a few permille and reduce the size of the effective couplings.

\begin{figure}
\centering \includegraphics{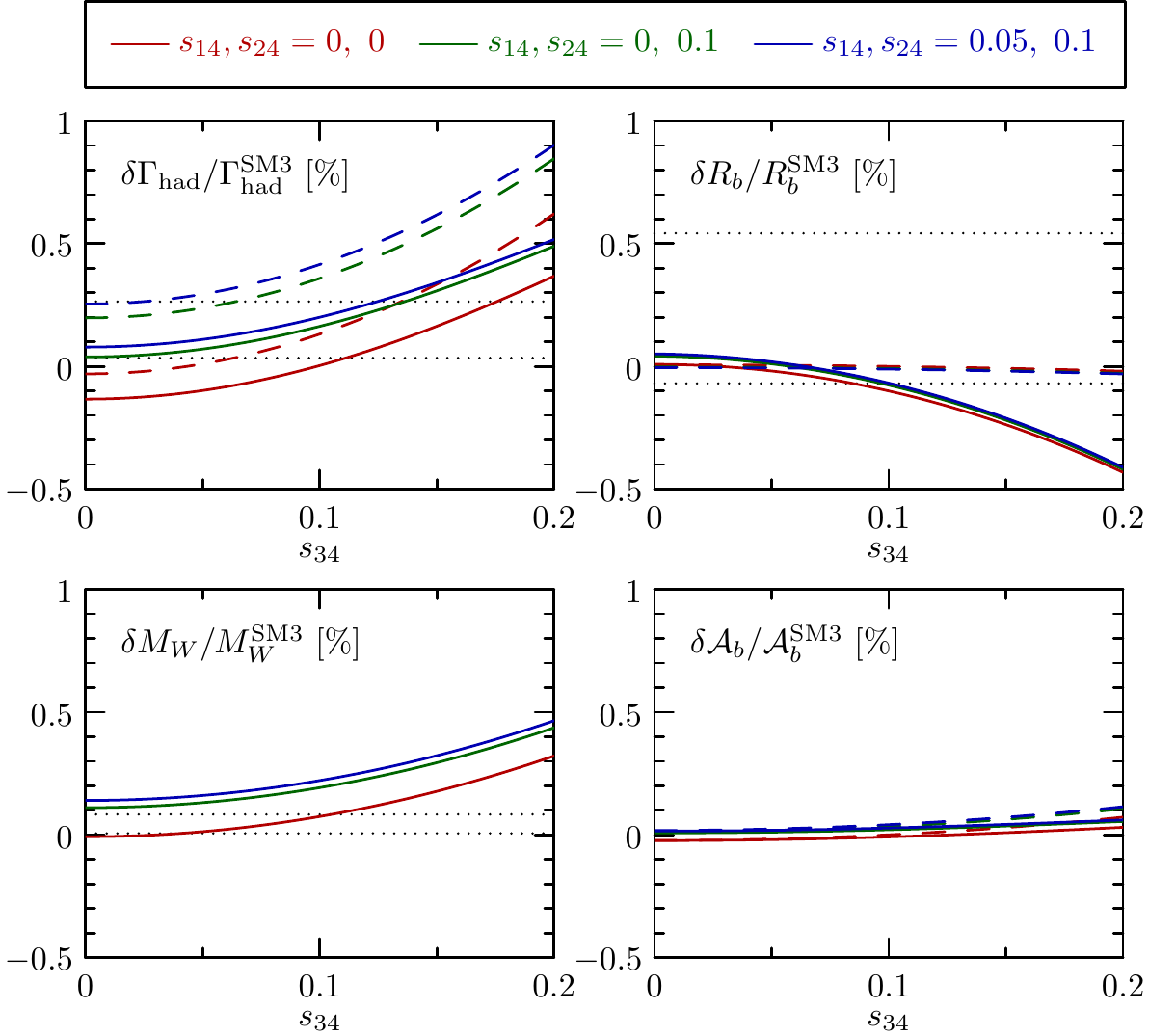}
\caption{The relative `new physics' corrections to the observables
  $\Gamma_\text{had}$, $R_b$, $M_W$ and $\Acal_b$ as functions of $s_{34}$ for
  the parameters \eqref{eq:first_scenario} and different combinations of
  $s_{14}$ and $s_{24}$. The dashed lines contain only the oblique
  corrections. The dotted lines indicate the range in which the observable would
  be in $1\sigma$ agreement with the experimental value. For $\Acal_b$ the
  dotted lines are outside the displayed range and all values are in $1\sigma$
  agreement with the measurement.}
\label{fig:s14s24}
\end{figure}
Figure~\ref{fig:s14s24} shows the relative new physics corrections to
$\Gamma_\text{had}$, $R_b$, $M_W$ and $\Acal_b$ as functions of $s_{34}$ for the
same combinations of $s_{14}$ and $s_{24}$. The corrections to
$\Gamma_\text{had}$ are between $-0.1$ and $0.5\%$. Varying $s_{24}$ in it's
allowed range shifts the ratio
$\delta\Gamma_\text{had}/\Gamma_\text{had}^\text{SM3}$ by approximately
$0.2\%$. The constraint imposed by $\Gamma_\text{had}$ on $\theta_{24}$ and
especially $\theta_{34}$ is therefore competitive with those obtained from
flavour physics. The oblique approximation over-estimates $\Gamma_\text{had}$ by
up to $0.4\%$. The hadronic branching fraction $R_b$ is essentially insensitive
to $\theta_{14}$ and $\theta_{24}$. The dependence on $\theta_{34}$ is not
captured at all by the oblique approximation and for $s_{34}=0.2$ it differs
from the exact result by $0.4\%$. The corrections to $M_W$ lie between 0 and
0.5\%. The corrections to $\Acal_b$ are of the order of one permille and thus
negligible compared to the experimental error on $\Acal_b$.

It is worth noting that agreement with the experimental value within
approximately one standard deviation can be achieved simultaneously for the two
most constraining observables, $\Gamma_\text{had}$ and $M_W$, even for nonzero
values of $s_{24}$ and $s_{34}$. In the scenario discussed above this happens,
for example, for $s_{24}=0$, $s_{34}=0.1$ or for $s_{24}=0.1$ and
$s_{34}=0$. This indicates that non-trivial mixing scenarios,
i.e. scenarios where all three fourth-generation mixing angles are nonzero, may
well be in good agreement with electroweak precision measurements and that
varying $\theta_{24}$ might even improve the electroweak fit. To quantify this
statement it would be necessary to perform a global fit of all SM4 parameters
(couplings, masses and CKM mixing angles) which takes into account electroweak
precision measurements and flavour observables at the same time. In such a fit
the non-oblique contributions to the EWPOs must be included since their effect
is comparable to the experimental error.

\begin{figure}
  \centering\includegraphics{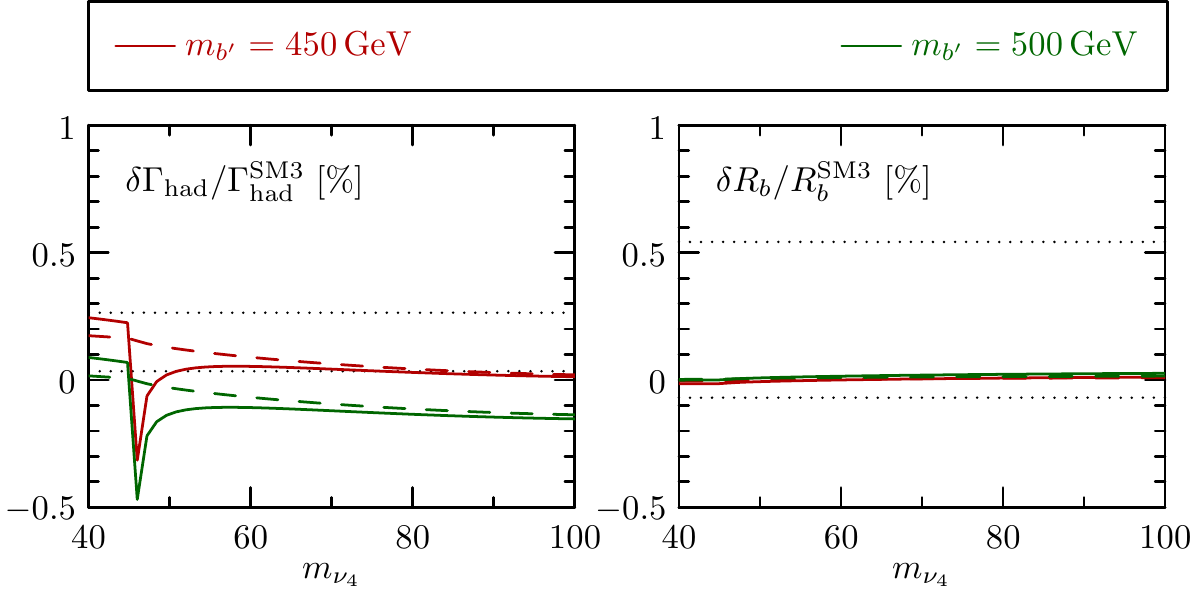}
  \caption{The dependence of $\Gamma_\text{had}$ and $R_b$ on $m_{\nu_4}$ for
    two different values of $m_{b'}$, $s_{14}=s_{24}=s_{34}=0$ and all other
    parameters according to \eqref{eq:first_scenario}. The dashed lines only
    contain the oblique corrections while the solid lines are the results of the
    exact calculation.}
  \label{fig:mn4}
\end{figure}
Let us finally discuss the quality of the oblique approximation as a function of
the fourth-generation neutrino mass. Figure~\ref{fig:mn4} shows the dependence
of $\Gamma_\text{had}$ and $R_b$ on $m_{\nu_4}$ for two different values of
$m_{b'}$, $s_{14}=s_{24}=s_{34}=0$ and all other parameters according to
\eqref{eq:first_scenario}. Again, the dashed lines only contain the oblique
corrections while the solid lines are the results of the exact calculation.  We
see that $R_b$ is insensitive to the neutrino mass and that (for vanishing
mixing angles) the oblique corrections agree with the exact results. For
$\Gamma_\text{had}$ the oblique effects become relevant for neutrino masses
below approximately \unit{60}{GeV}. At the threshold $m_{\nu_4}=M_Z/2$ the
full one-loop results diverge and become unreliable.
% 
%
%
%%%%%%%%%%%%%%%%%%%%%%%%%%%%%%%%%%%%%%%%%%%%%%%%%%%%%%%%%%%%%%%%%%%%%%%%%%%%%%%%
\section{Conclusion}\label{sec:conclusions}
%%%%%%%%%%%%%%%%%%%%%%%%%%%%%%%%%%%%%%%%%%%%%%%%%%%%%%%%%%%%%%%%%%%%%%%%%%%%%%%%
%
%
In this paper we studied the potential size of the corrections to the EWPOs that
arise in a fourth-generation model with the most general (quark-)flavour
structure (but assuming a trivial flavour-structure in the lepton-sector).

By computing ratios of form factors in the SM4 and SM3 to one-loop order and
scaling results of precision calculations for the corresponding SM3 form factors
with these ratios we can compute EWPOs within the SM4, including factorisable
higher-order QCD and QED corrections and non-oblique contributions. The ratios
of form factors can easily be computed with automated tools for one-loop
calculations such as \code{FeynArts/FormCalc} \cite{Hahn:2006qw, Hahn:2000kx,
  Hahn:1998yk}. This way, no approximations beyond the one-loop approximation
are made for the new-physics contributions. In particular, non-obliqe
contributions are included correctly this way. The application of our method
to other new-physics models is straightforward.

We find that the EWPOs are sensitive to the fourth-generation mixing angles
$\theta_{24}$ and $\theta_{34}$. The constraints imposed by $\Gamma_\text{had}$
and $M_W$ are generally as strong as those obtained from flavour physics. In the
region of parameter space that is currently favoured by direct searches for
Higgs bosons and fourth-generation quarks the non-oblique corrections to the
electroweak observables are of the order of a few permille and should therefore
be included in a global fit.

\paragraph*{Acknowledgement.}

The authors would like to thank Alexander Lenz and Ulrich Nierste for fruitful
discussions and thorough proof reading.

J.R. is supported by DFG Sonderforschungsbereich SFB/TR 9 ``Computergest\"utzte
Theoretische Teilchenphysik''. P.G. is supported by DFG SFB/TR9.
M.W. is partially supported by project DFG NI 1105/2-1.

\begin{appendix}
%
%
%
%%%%%%%%%%%%%%%%%%%%%%%%%%%%%%%%%%%%%%%%%%%%%%%%%%%%%%%%%%%%%%%%%%%%%%%%%%%%%%%%
\section{Parametrisation of the CKM Matrix}\label{sec:CKM}
%%%%%%%%%%%%%%%%%%%%%%%%%%%%%%%%%%%%%%%%%%%%%%%%%%%%%%%%%%%%%%%%%%%%%%%%%%%%%%%%
%
%
The CKM matrix of the SM3 depends on three mixing angles $\theta_{12}$,
$\theta_{13}$ and $\theta_{23}$ and one complex phase $\delta_{13}$.
Its elements are given by
\begin{equation}\label{eq:VCKM3}
  V_\text{CKM} = \begin{pmatrix}
      c_{12} c_{13}
    & s_{12} c_{13}
    & s_{13} e^{-i\delta_{13}}
    \\
      -s_{12} c_{23} - c_{12} s_{23} s_{13} e^{i\delta_{13}}
    & c_{12} c_{23} - s_{12} s_{23} s_{13} e^{i\delta_{13}}
    & s_{23} c_{13}
    \\
      s_{12} s_{23} - c_{12} c_{23} s_{13} e^{i\delta_{13}}
    & -c_{12} s_{23} - s_{12} c_{23} s_{13} e^{i\delta_{13}}
    & c_{23} c_{13}
  \end{pmatrix}
  \eqsep,
\end{equation}
where $c_{ij}=\cos\theta_{ij}$ and $s_{ij}=\sin\theta_{ij}$. The SM4 CKM matrix
is parametrised by three additional mixing angles $\theta_{14}$, $\theta_{24}$
and $\theta_{34}$ and two additional phases $\delta_{14}$ and $\delta_{24}$. In
terms of these parameters, it is then written as
\begin{equation}
  V_\text{CKM4} = \begin{pmatrix}
    V_{ud} & V_{us} & V_{ub} & V_{ub'} \\
    V_{cd} & V_{cs} & V_{cb} & V_{cb'} \\
    V_{td} & V_{ts} & V_{tb} & V_{tb'} \\
    V_{t'd} & V_{t's} & V_{t'b} & V_{t'b'}
  \end{pmatrix}
\end{equation}
with
\begin{gather}
  V_{ud} = c_{12} c_{13} c_{14}\eqsep,\eqsep
  V_{us} = c_{13} c_{14} s_{12}\eqsep,\eqsep
  V_{ub} = c_{14} s_{13}e^{-i\delta_{13}}\eqsep,
  \nonumber\\
  V_{ub'} = s_{14}e^{-i\delta_{14}}\eqsep,\eqsep
  V_{cb'} = c_{14} s_{24}e^{-i\delta_{24}}\eqsep,\eqsep
  V_{tb'} = c_{14} c_{24} s_{34}\eqsep,\eqsep
  V_{t'b'} = c_{14} c_{24} c_{34}\eqsep,
  \nonumber\displaybreak[0]\\
  \begin{aligned}
    V_{cd} &= -c_{23} c_{24} s_{12}
      + c_{12} (-c_{24} s_{13} s_{23} e^{i\delta_{13}}
         - c_{13} s_{14} s_{24}e^{i(\delta_{14}-\delta_{24})})\eqsep,\\
    V_{cs} &= c_{12} c_{23} c_{24}
      + s_{12} (-c_{24} s_{13} s_{23} e^{i\delta_{13}}
         - c_{13} s_{14} s_{24}e^{i(\delta_{14}-\delta_{24})})\eqsep,\\
    V_{cb} &= c_{13} c_{24} s_{23}
      - s_{13} s_{14} s_{24}e^{i(\delta_{14}-\delta_{13}-\delta_{24})}\eqsep,
    \\
    V_{td} &= -s_{12} (-c_{34} s_{23} - c_{23} s_{24} s_{34} e^{i\delta_{24}})\\
    &\phantom{{}={}} + c_{12} (-c_{13} c_{24} s_{14} s_{34} e^{i\delta_{14}}
      -s_{13}e^{i\delta_{13}}
        (c_{23} c_{34} - s_{23} s_{24} s_{34} e^{i\delta_{24}}))\eqsep,\\
    V_{ts} &= c_{12} (-c_{34} s_{23} - c_{23} s_{24} s_{34} e^{i\delta_{24}})\\
    &\phantom{{}={}} + s_{12} (-c_{13} c_{24} s_{14} s_{34} e^{i\delta_{14}}
      - s_{13}e^{i\delta_{13}} (c_{23} c_{34}
           - s_{23} s_{24} s_{34} e^{i\delta_{24}}))\eqsep,\\
    V_{tb} &= -c_{24} s_{13} s_{14} s_{34} e^{i(\delta_{14}-\delta_{13})}
      + c_{13} (c_{23} c_{34} - s_{23} s_{24} s_{34} e^{i\delta_{24}})\eqsep,
  \\
    V_{t'd} &= -s_{12} (-c_{23} c_{34} s_{24}e^{i\delta_{24}} + s_{23} s_{34})\\
    &\phantom{{}={}} + c_{12} (-c_{13} c_{24} c_{34} s_{14}e^{i\delta_{14}}
      - s_{13}e^{i\delta_{13}} (-c_{34} s_{23} s_{24}e^{i\delta_{24}}
        - c_{23} s_{34}))\eqsep,\\
    V_{t's} &= c_{12} (-c_{23} c_{34} s_{24}e^{i\delta_{24}} + s_{23} s_{34})\\
    &\phantom{{}={}} + s_{12} (-c_{13} c_{24} c_{34} s_{14}e^{i\delta_{14}}
      - s_{13}e^{i\delta_{13}} (-c_{34} s_{23} s_{24}e^{i\delta_{24}}
        - c_{23} s_{34}))\eqsep,\\
    V_{t'b} &= -c_{24} c_{34} s_{13} s_{14} e^{i(\delta_{14}-\delta_{13})}
      + c_{13} (-c_{34} s_{23} s_{24}e^{i\delta_{24}} - c_{23} s_{34})\eqsep.
  \end{aligned}
\end{gather}
For $\theta_{14}=\theta_{24}=\theta_{34}=\delta_{14}=\delta_{24}=0$ the
SM4 CKM matrix assumes block-diagonal form with the SM3 CKM matrix in the first
$3\times 3$ block:
\begin{equation}
  V_\text{CKM4} = \begin{pmatrix}
      &              &   & 0 \\
      & V_\text{CKM} &   & 0 \\
      &              &   & 0 \\
    0 & 0            & 0 & 1
  \end{pmatrix}
  \eqsep.
\end{equation}

\end{appendix}

\bibliography{ewpsm4}

\end{document}